\documentclass{article}
\usepackage{amsmath}
\usepackage{graphicx}
\usepackage{authblk}
\numberwithin{equation}{section}
\begin{document}

\title{Search for Magnetic Monopole using ICAL at INO}
\author[1,2,$\dagger$]{ N. Dash}
\author[1,2]{V. M. Datar\thanks{vivek.datar@gmail.com}}
\author[3]{ G. Majumder}

\affil[1]{Nuclear Physics Division, Bhabha Atomic Research Centre, Mumbai - 400085, INDIA}
\affil[2]{Homi Bhabha National Institute, Anushaktinagar, Mumbai - 400094, INDIA}
\affil[3]{Tata Institute of Fundamental Research, Mumbai - 400005, INDIA}
\affil[$\dagger$]{India-based Neutrino Observatory}

\maketitle

\begin{abstract}
Sub-relativistic magnetic monopoles are predicted from the GUT era by theory. To date there have been no confirmed observations of such exotic particles. The Iron CALorimeter (ICAL) at India-based Neutrino Observatory (INO) aims to measure the neutrino oscillation parameters precisely. As it is a tracking detector there is also the possibility of detecting magnetic monopoles in the sub-relativistic region. Using ICAL the magnetic monopole event is characterised by the large time intervals of upto $30~\mu sec$ between the signals in successive layers of the active detectors. The aim of this study is to identify the sensitivity of ICAL for a particle carrying magnetic charge in the mass range from $10^{5}~-~10^{17}$~GeV with $\beta$ ranging from $10^{-5}~-~9\times10^{-1}$ for ICAL at INO. A similar study has also been carried out for the ICAL prototype which will be placed overground. Due to the rock cover of approximately $1.3~km$, ICAL at INO will not be able to place bounds on the flux of the lower mass magnetic monopoles. This mass region is however addressed by the prototype ICAL.
\end{abstract}


\section{Introduction}
The India-based Neutrino Observatory (INO) $\bf [1]$ project is a proposed multi-institutional effort aimed at building an underground laboratory with a rock cover of approximately $1.3~km$ at Pottipuram in Bodi West hills of Theni District of Tamil Nadu. It will focus on research in the area of non-accelerator based high energy physics and nuclear physics. The flagship experiment, to study atmospheric neutrinos, will make use of a large detector called the Iron CALorimeter (ICAL). It is a magnetised calorimeter of mass around $50~kton$. The detector size is of the order of $48~m~\times~16~m~\times~15~m$. Due to its comparatively large size, assuming an isotropic $\bf [2]$ magnetic monopole (MM) flux equal to the Parker Limit $\bf [3]$, ICAL could expect a few events due to MMs per year. The detectors which are used for detecting MM are mainly based either on induction or on ionization. During the early 80's, the experiment at Stanford University by Cabrera $\bf [4]$ performed a mass and velocity independent search of MM by the induction method using a Superconducting QUantum Interferometer Device (SQUID) as the detector. The induction based experiment was subsequently improved upon to yield the upper bound of the flux of MMs $\sim$ $3.8\times10^{-13}~cm^{-2}~sr^{-1}~sec^{-1}$ $\bf [5]$. The experiments like MACRO $\bf [6]$, SLIM $\bf [7]$, Soudan 2 $\bf [8]$ etc., used the ionization method for MM detection as they have used either scintillators or gaseous detectors. Similarly most of the ice and water Cherenkov based detectors like AMANDA $\bf [9]$, Baikal $\bf [10]$, Kamiokande $\bf [11]$ etc. have also looked for MM detection and have placed upper bounds on MM flux. Last, but not least, (the accelerator based ) experiments such as CDF $\bf [12]$, Oklahoma $\bf [13]$ etc., have looked for low mass MM which may be produced at accelerators. The ICAL detector size is comparable to the size of the MACRO detector at Gran Sasso. The ionization produced in a gas detector can either be measured quantitatively with a signal height ``proportional to'' energy loss of the MMs in the gas thickness or produce a saturated pulse which only carries ``hit'' and ``time'' information. For an MM, in ionization method the energy loss in the active detector element should be sufficient enough to give a signal in the detector. The ICAL will use a gaseous detector called the Resistive Plate Chamber (RPC), which belongs to the second category of gas detector mentioned earlier. The track of the MM in the ICAL and the characteristic sequence of trigger times of consecutive layers of RPCs will help in identifying the MM against random background.

In this work we report the results of the simulation work that has been carried for MMs with Dirac charge $(g=68.5e)$ $\bf [14]$ to estimate the upper limit in flux for ICAL within a mass range of $10^5 - 10^{18}$~GeV and velocity range of $10^{-5}$c to $0.9$c. A prototype of the ICAL with a mass of $\sim 2~kton$ is planned to be built at the surface. This allows it to be sensitive to lighter MMs. A similar analysis has been carried out for the ICAL prototype to estimate the efficient region in the MM M$-\beta$ plane. In Sec.2, we briefly discuss the physics of Magnetic Monopoles. In Sec.3, we outline the mechanisms by which an MM gives a signal in the detector and makes it detectable like a particle carrying electric charge. In Secs.4 and 5, we discuss the future prospect of the MM in the ICAL. Section 6, gives a brief idea about the possibility of using the ICAL prototype for detecting MMs and the paper ends  with conclusions in the last section.


\section{Theoretical Overview}
In 1931 Dirac predicted the possible existence of isolated magnetic poles, by considering a connection between a charge on the MM and the quantization (integral multiple of smallest unit) $\bf [14]$ of an electric charge, given by
\begin{equation}\label{eq:someequation}
eg = \frac{n\hbar c}{2\pi},
\end{equation}
where $g$ is the magnetic charge, $e$ is the electric charge, and $n$ is an integer which can take both positive and negative values.

The expression in Eq.(2.1) was also derived by M. N. Saha $\bf [15]$ in $1936$, by considering the quantized angular momentum perpendicular to the line joining the point electric charge and MM.

In 1974 G. 't Hooft $\bf [16]$ and Polyakov $\bf [17]$ discovered MM solutions of the classical equations of motion for spontaneously broken non-abelian gauge field theories. This leads to a lower bound on the mass of the MM which depends on the mass of the carrier of the unified interaction and coupling constant.
\begin{equation}
M_{mm} \geq \frac{M_x}{G},
\end{equation}
where $M_x$ is the mass of the carrier of the unified interaction and $G$ is the unified coupling constant.

 From a cosmological point of view it is believed that MMs are created during the big-bang around $10^{-34}~sec$ after of the creation of the universe. It is assumed that they are moving with velocity of the earth$^{'}$s orbital velocity of the order of $10^{-3}$~c. Those with large mass can penetrate the rock with an initial velocity of the order of the orbital velocity of the earth. Due to the energy loss of MMs in the rock their velocity gradually decreases. If the mass of the MM is small they get stopped in the earth matter. Parker $\bf [3]$ has obtained an upper bound on the flux of the MMs in the galaxy by noting that the rate of energy loss is small compared to the time scale on which the galactic field regenerated. The flux for the MM with mass less than equal to $10^{17}$~GeV is constant but it increases linearly with mass for higher masses (see Eq.(2.3)).
\begin{equation}
F_M = \begin{cases} {10^{-15}cm^{-2}sec^{-1}sr^{-1}}, & {M \leq 10^{17}GeV}\\{10^{-15}\frac{M}{10^{17}GeV}cm^{-2}sec^{-1}sr^{-1}}, & {M \geq 10^{17}GeV} \end{cases}
\end{equation}
 
The existence of magnetic charges and magnetic currents make Maxwell's equations symmetric. The Maxwell's equation in the cgs system of units are :
\begin{subequations}
\begin{align}
\nabla~.~E & = 4\pi\rho_e\\ 
\nabla~.~B & = 4\pi{\rho_m} \\
\nabla \times E & = - \frac{1}{c}\frac{\partial B}{\partial t} - \frac{4\pi}{c}{J_m}\\
\nabla \times B & = \frac{1}{c}\frac{\partial E}{\partial t} + \frac{4\pi}{c}J_e
\end{align}
\end{subequations}
where $J_e$ is the electric current density, $\rho_e$ is the electric charge density, $\rho_m$ is the magnetic charge density, and $J_m$ is the magnetic current density.


\section{Interaction of a magnetic monopole with matter}
The detection of an MM in a particle detector is similar to that of a charged particle as both are detected through a certain amount of energy loss in the detector medium. Just as the stopping power of a charged particle depends on its charge and velocity, the energy loss of an MM also depends on its magnetic charge and velocity.

A fast MM is characterized by its velocity [$v=\beta c$], which is much larger than the typical electron orbital 
velocity ($\alpha c$, $\alpha$ being the fine structure constant). So in principle, for a fast MM, the electron can be assumed to be stationary so that the MM-e collisions can be characterized by an impact parameter. In close collisions, the large energy transfer from the projectile allows the electrons to be treated as free. But in the case of distant collisions the atoms are excited by the perturbation caused by electric field of the MM. These approximations are valid for a particle irrespective of whether it carries an electric or magnetic charge. However there is a difference between the dependence of the stopping power of fast MM and an electric charge through the dependence on velocity. For an MM, the magnitude of the electric field is reduced by a factor of $\beta$ from that of an electrically charged particle with the same velocity and charge. For $\beta \geq \alpha$ it is essentially the lab frame electric field which determines the interaction between the projectile and the electrons. Since the stopping power scales as the square of the electric field, it is apparent that the ratio of fast MM stopping power to that of an electric projectile is $\sim$ $({g\beta}/{Ze})^2$. Hence the stopping power expression is similar to that of an electrically charged particle which is given by 
Bethe-Bloch expression $\bf [18]$. In this expression the $Ze$ term is replaced by $g\beta$ for the MM is given by 
\begin{equation}
-\frac{dE}{dX}=\frac{4\pi {N{e}}^{2}g^{2}}{m_{e}c^{2}}(\ln(\frac{2m_{e}c^{2}\beta^{2}\gamma^{2}}{I_{m}})-\frac{1}{2}-\frac{\delta_{m}}{2} 
-B(\mid g \mid)+\frac{K(\mid g \mid)}{2}),
\end{equation}
where $N$ is the electron number density, $m_e$ is the mass of the electron, $I_m$ is the mean ionization potential of the target material, $\delta_{m}$ is the density correction for high energies, 
\begin{equation*}
B(\mid g \mid) =\begin{cases}{0.248} & {\mid g \mid=137e/2}\\
 {0.672} & {\mid g \mid=137e} \end{cases}
\end{equation*}
is the Bloch correction for an MM, and 
 
 \begin{equation*}
 K(\mid g \mid) = \begin{cases}{ 0.406} & {\mid g \mid=137e/2}\\
 {0.346} & {\mid g \mid=137e} \end{cases}
\end{equation*}
is the Kazama cross-section correction $\bf [19]$.
At the above mentioned velocity range the ratio between the stopping power of the MM and the relativistic singly charged electric particle is $\sim$ $(g/e)^2$ $\sim$ $4700$. Due to its enormous energy loss, an MM can be detected easily by any device that can detect electrically charged particles.

The expression in Eq.(3.1) is valid only for high values of $\beta$, viz. $\beta \geq 10^{-2}$. However for lower $\beta$, the interaction between the MM and the absorber is different. For an absorber with atoms with larger atomic numbers [$Z \geq 10$] the electron motion in an atom is approximated by that in a locally uniform potential, such as obtained from a Thomas-Fermi approximation. But for lower Z materials the excitation of the atom is due to the interaction of the magnetic field of the MM with the electron magnetic moment, where the energy loss varies linearly with $\beta$ $\bf [20]$. The expression is as follows
\begin{equation}
-\frac{dE}{dX}=20(\frac{g}{137e})^2\beta~\frac{GeV}{gm~cm^{-2}}.
\end{equation}
The energy loss of MMs at lower $\beta$ is comparatively less than that for electrically charged particles. This can be used to distinguish MM from electrically charged particles $\bf [21]$.

Figure 1 shows the stopping power of an MM with Dirac charge in silicon computed using GEANT4, the High Energy Physics (HEP) simulation tool-kit $\bf [22]$ and is comparable with the plot presented in Ref. $\bf [18]$.
 
\begin{figure}[here]
\center
\includegraphics[width=66mm]{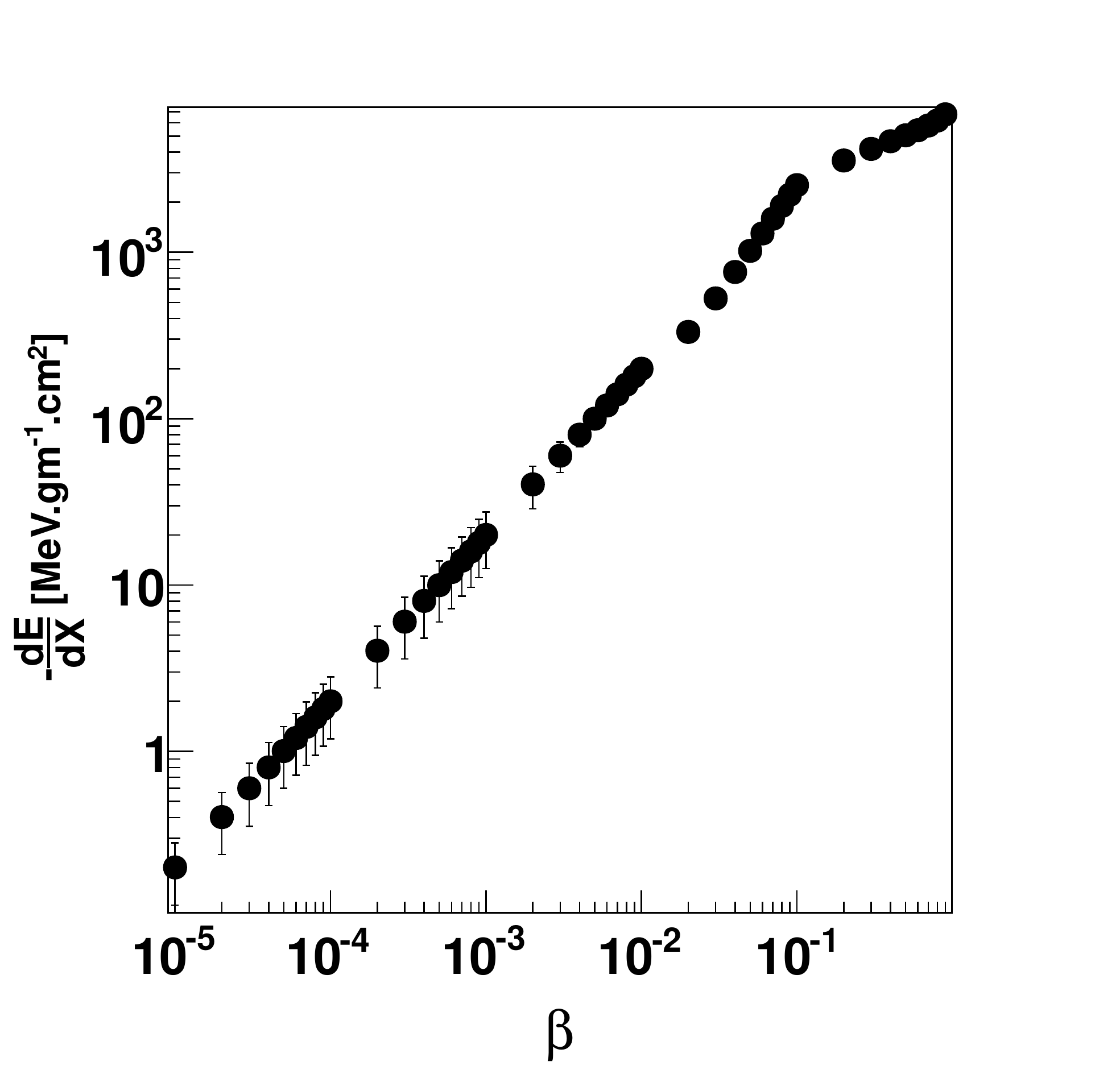}
\caption{\label{fig1}Stopping power of an MM($g=68.5e$) in silicon.}
\end{figure}


\section{Magnetic Monopole simulation for ICAL}
GEANT4 is used to define the ICAL detector geometry as well as to simulate the detector response for an MM event, through the simulated detector volume. The detector design which has been implemented in GEANT4 consists of $3$ ICAL modules each of $16~kton$. Each ICAL module consists of $150$ layers of $2~m~\times~4~m~$ iron plates of $5.6~cm$ thickness interleaved with layers of RPCs. The iron plates are magnetised with average field of $1.3~Tesla$. The transverse dimensions of each module are $16~m~\times~16~m$. Two consecutive iron plates are separated by an air gap of $4~cm$ to accommodate a layer of RPCs which are the active detector elements. The RPC consists of a $2~mm$ layer of gas enclosed between $3~mm$ thick glass sheets, which are overlaid by honey-comb based pick-up strips for electrical contact which yield an electrical signal on the passage of a charged particle or an MM. The pick-up readout system on either side of the RPC are placed orthogonal to each other to give both X and Y co-ordinate information. Each strip has a width of $\sim$ $3~cm$. The RPC also gives a fast timing signal ($\sigma$~$\sim$~$1~nsec$) that can be used to discriminate MMs from muons. The ICAL modules have the same dimensions and are placed side by side, separated by a $40~cm$ gap.

To simulate the MM events for the ICAL at INO, a rock mass of density $2.89$~${gm}/{cm^3}$ of height $1.3~km$ from the top surface of the detector is defined in addition to the ICAL geometry (see Fig. 2). Particles are incident from the surface of the rock, so that they will move through the rock before detection in ICAL and are represented by solid line in Fig. 2. The events presented by dashed line in Fig. 2 are not used during simulation. To simulate an isotropic flux the zenith angle (cos$\theta$) is smeared from $\pi$ to ${\pi}/{2}$ corresponding to down-ward going events and $2\pi$ smearing in azimuthal angle ($\phi$). The stopping power in different materials is given by Eqs.(3.1 and 3.2). 
\begin{figure}[here]
\center
\includegraphics[width=76mm]{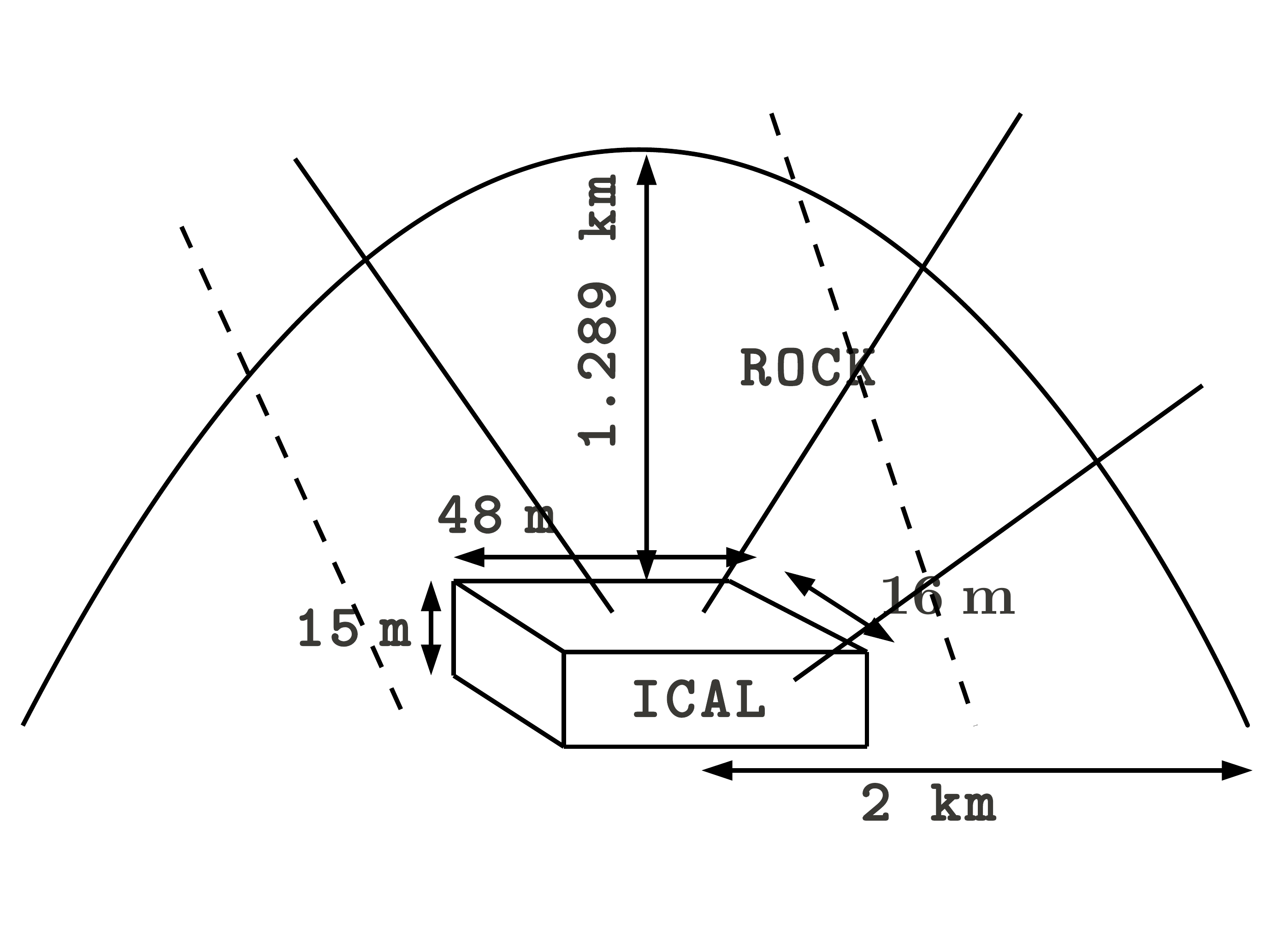}
\caption{\label{fig1}Schematic view of the MM events generation in the ICAL detector with rock.}
\end{figure}

Particles propagate through the detector while losing energy in the sensitive region of the detector which has a gas mixture of Freon (95.15$\%$), Iso-Butane (4.51$\%$) and $SF_6$ (0.34$\%$), and registering ``hits''. Each hit has an associated position and time information. The plot in Fig. 3 shows the energy loss of an MM in $2~mm$ thick RPC gas with composition mentioned above as a function of $\beta$ range. At lower $\beta$ it increases linearly with it but at higher value it increases ln($\gamma^2$).
\begin{figure}[here]
\center
\includegraphics[width=66mm]{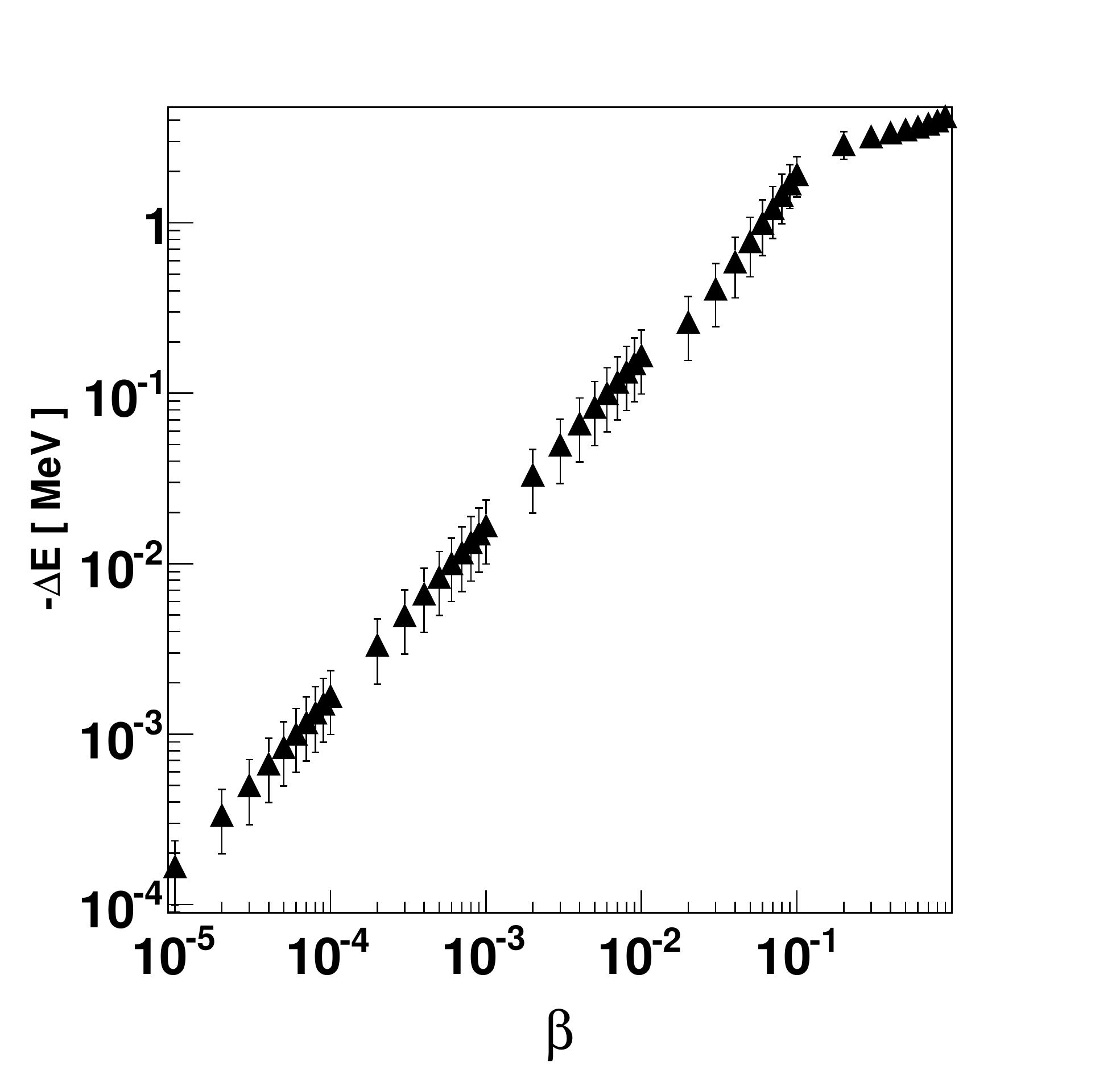}
\caption{\label{fig1}Energy Loss of an MM in 2 mm thick RPC gas(composition of gas is in text) as function of velocity in units of c.}
\end{figure}


\section{Analysis and Results}
The time and position information of each hit is used to reconstruct the velocity of a particle. Their massive nature do not allow them to bend in the $1.3~Tesla$ magnetic field in the iron plates during their travel through the detector. So its trajectory is almost straight. Hence a straight line fitting is used for velocity reconstruction. Due to the large mass and sub-relativistic velocity, the time of flight method is suitable for identifying the MM using ICAL. For relativistic MMs high energy muons will constitute the main background. However in the smaller velocity region the background will be due to the chance coincidences rate which can be minimised by choosing a minimum number of layers for velocity reconstruction. By requiring a coincidence of an additional layer, the random coincidences reduce by a factor of $\sim~2\times10^{-6}$ for a $\beta$ value of $0.8$ with a travel distance of $10~cm$ and assuming RPC strip rate to be $200~Hz$. For each mass and $\beta$ bin a sample of 10,000 events have been used to estimate the efficiency.
\begin{figure}[here]
\center
\includegraphics[width=66mm]{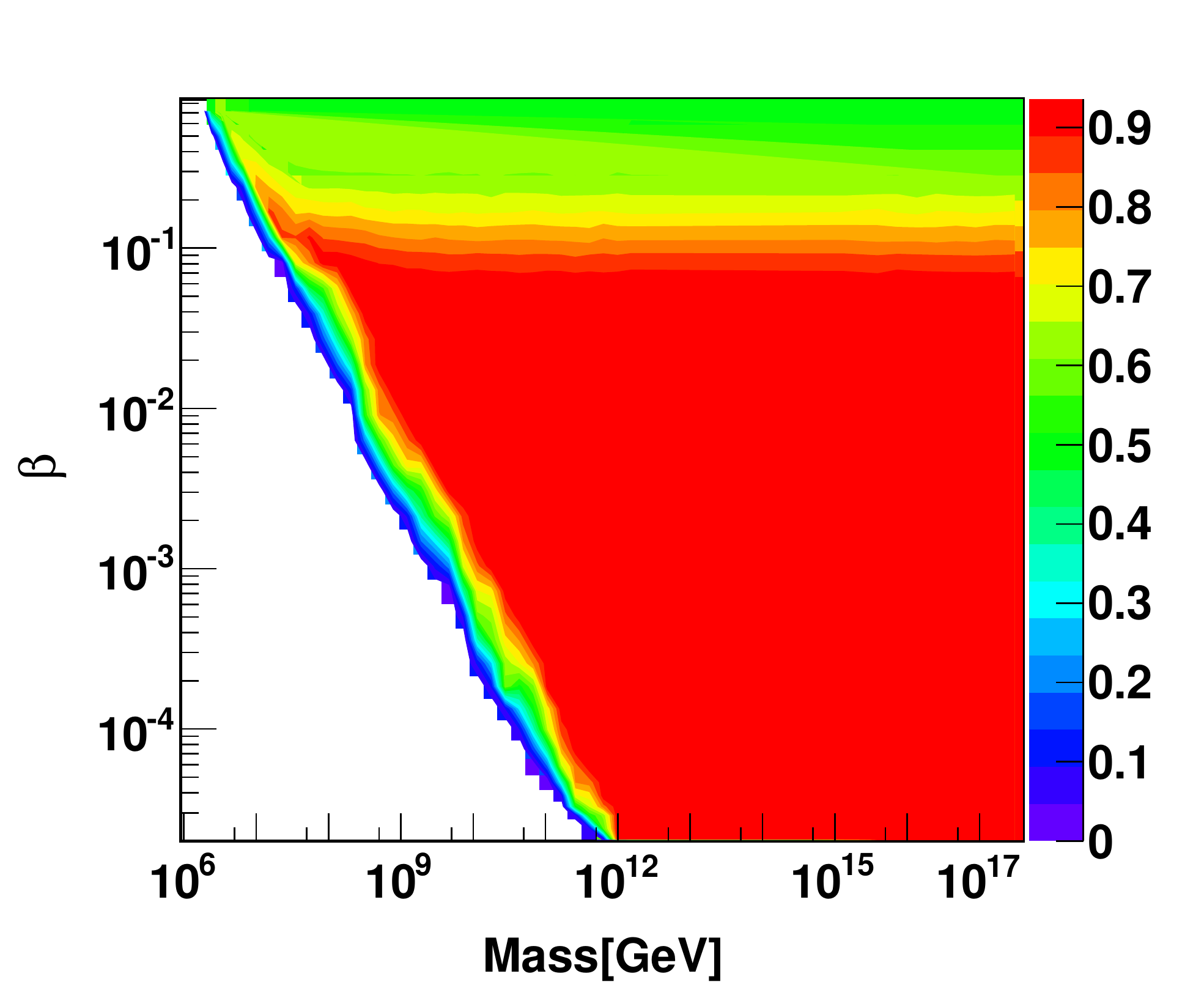}
\caption{\label{fig1}ICAL detection efficiency for MM in its M$-\beta$ plane. The efficiencies in different regions in the plot are marked by different colours as shown in legend on the right side.}
\end{figure}

Figure 4 represents the sensitive region for ICAL in the M$-\beta$ plane by considering 10 as the minimum number of layers for velocity reconstruction. The events with reconstructed $\beta$ within $3$ times the incident $\beta$ are considered for the efficiency calculation. The different colours in the plot depict the efficiency in different regions of the M$-\beta$ plane with the value shown by the legend on the right side of the plot. In the plot the uncoloured part represents there is no sensitivity of ICAL, red colour represents detection efficiency with $90\%$ and colour coded such that the decreasing efficiency is depicted by colour with decreasing wavelength. At the lower mass and sub-relativistic $\beta$, they get absorbed by the rock materials before detecting in the ICAL. Similarly for higher mass and lower $\beta$ their energy is not sufficient to detect them by the active detector element of ICAL. In the mass range from $10^{12}$ to $10^{17}$~GeV, even at lower values of $\beta$ the MMs are able to travel at least 10 layers.

Depending on the minimum number of layers used for $\beta$ reconstruction the angular distributions obtained by the ICAL are different. The efficiency of detection of the MM by ICAL is used to then calculate the expected event rate or obtain a limit on the flux of MM from an upper bound on the detected events. Figure 5 shows the angular distribution for different minimum number of layers with different colours as marked inside it. The figure shows that with increase in cut in number of minimum layers the events are biased to more vertical ones to the detector and gradually the number of reconstructed events also goes on decreasing. So when the minimum layers are $10$ and $20$ all of them come from the upper half of the hemisphere. 

\begin{figure}[here]
\center
\includegraphics[width=66mm]{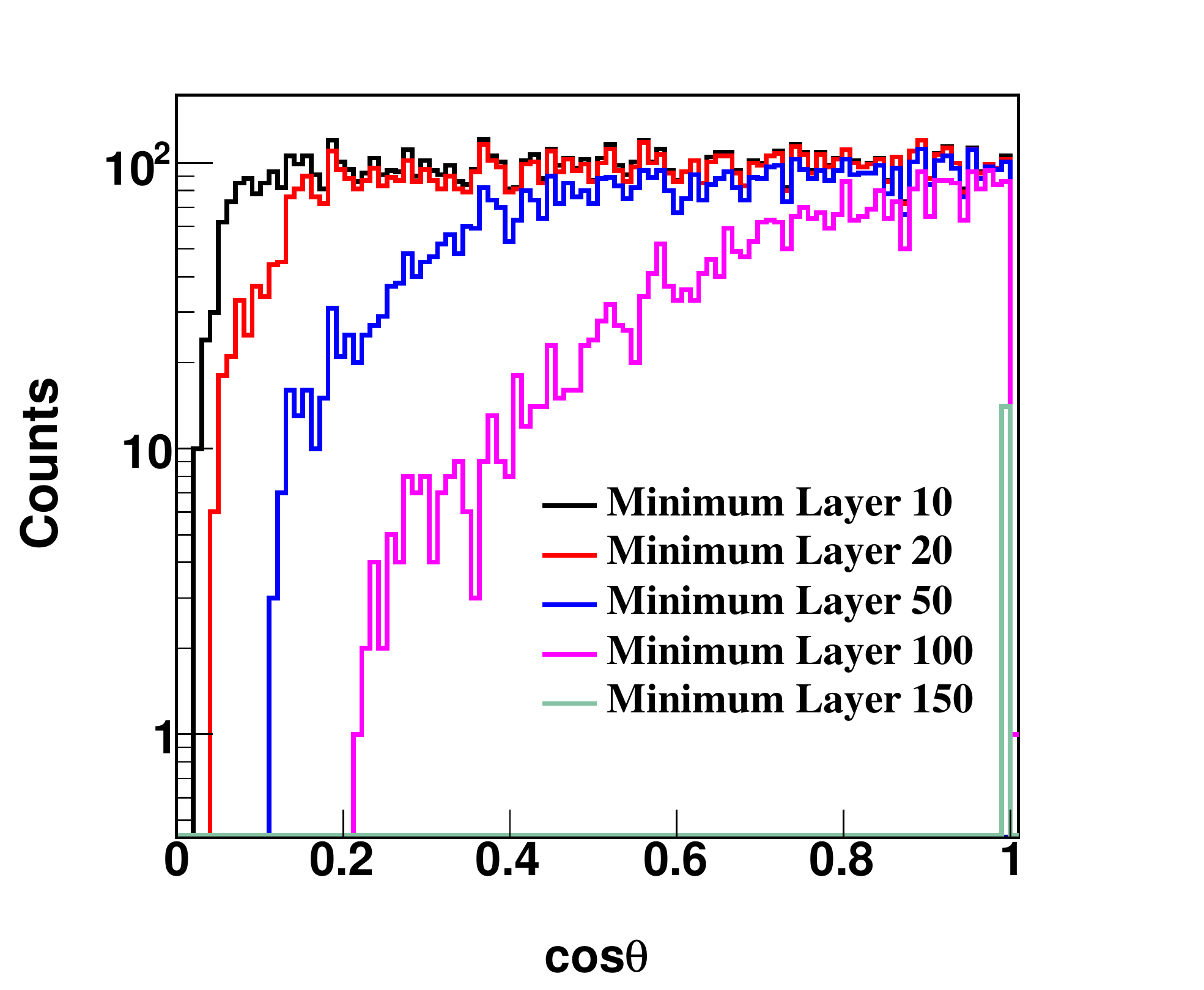}
\caption{\label{fig1}Reconstructed cos$\theta$ distribution for MM by considering minimum number of layers as 10, 20, 50, 100 and 150 for $\beta$ reconstruction in ICAL.}
\end{figure}
\subsection{Expected Event Rate}
If the effective area of cross-section of the ICAL detector is A, the solid angle obtained by its is $\Omega$, detection efficiency of MM estimated by ICAL is $\epsilon$, MM flux is f and T is the counting time period, then the expected events ($N_{Ex}$) is given by
\begin{equation}
N_{Ex}=f(cm^{-2}~sr^{-1}~sec^{-1})~A(cm^2)~\Omega(sr)~T(sec)~\epsilon.
\end{equation}
Furthermore, if we choose the ICAL transverse area of cross-section A = $16~m~\times~48~m~$ = $768$~$m^2$, $\Omega$ = $2\pi$ sr, T = $1$~Yr, $\epsilon$ = $1$ and f = $10^{-15}$ $cm^{-2}sr^{-1}sec^{-1}$, we get $N_{Ex}$ $\simeq$ $1.5$ events per year.
\begin{figure}[here]
\center
\includegraphics[width=66mm]{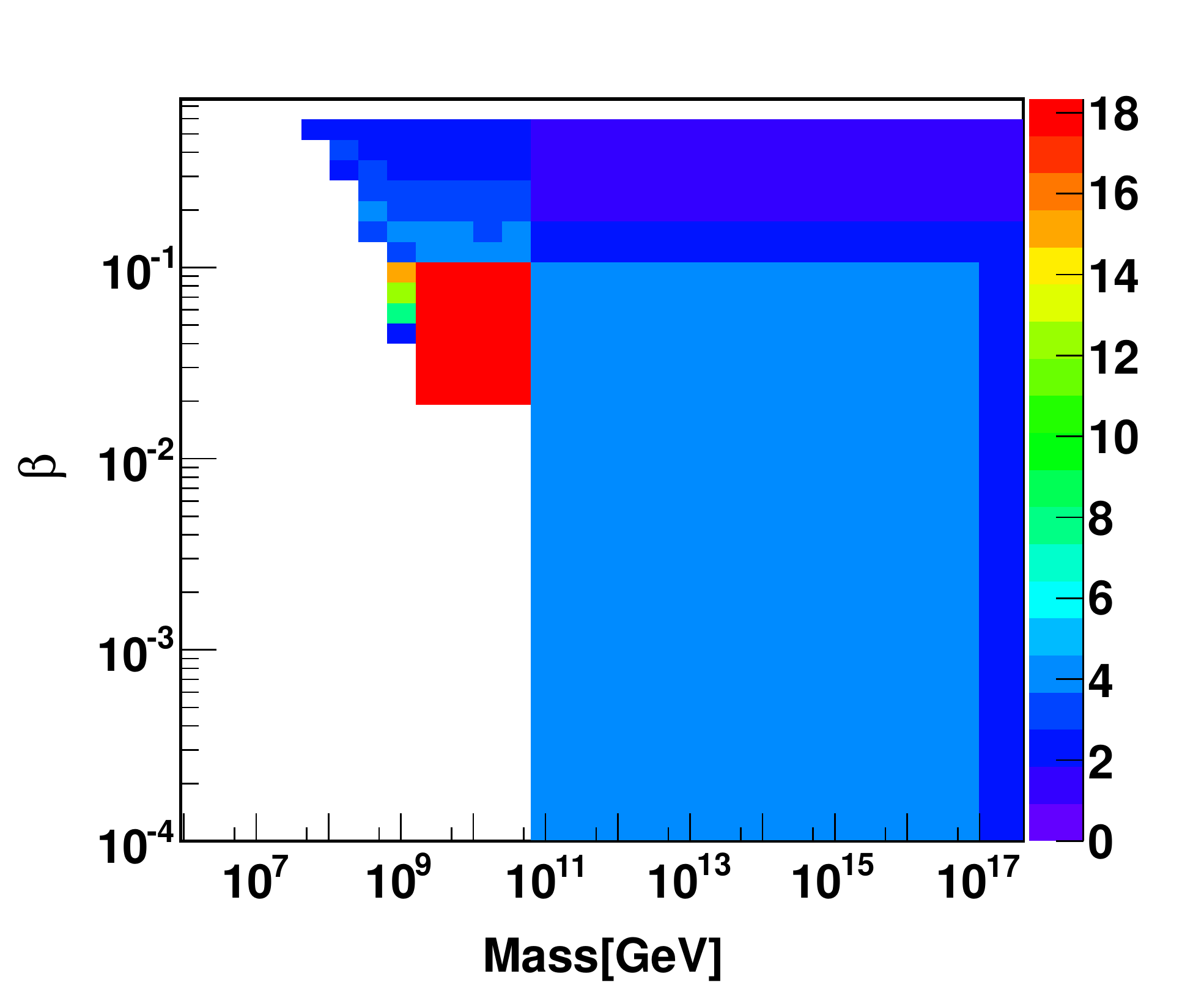}
\caption{\label{fig1}Expected events obtained for ICAL in 10 years of counting period using flux upper bound from the MACRO and SLIM experiments.}
\end{figure}

The expected events are guided by the upper limit in flux observed by different experiments. Figure 6 shows expected number of events estimated for 10 years of running using the flux upper bound obtained from the MACRO $\bf [6]$ and SLIM $\bf [7]$ experiments. The transverse area of cross-section of the ICAL, detection efficiency for MM from Fig. 4 and solid angle obtained by the ICAL for minimum 10 layers from Fig. 5 are used for the estimation of expected events.
\subsection{Estimation of Upper Limit in Flux}
An estimation of the Upper Limit in flux that can be obtained by ICAL assuming zero observed event is based on the Frequentist $\bf [23]$ Method including background. If the upper limit in number of event for finite observed event is $N_{upper}$, $N_{obs}$ is the number of observed events and $N_{BG}$ is the number of background events, then the upper limit in flux is given by 
\begin{equation}
f_{upper}=\frac{N_{upper}(N_{obs},N_{BG})}{A(cm^2)~\Omega(sr)~T(sec)~\epsilon}.
\end{equation}
For zero observed events and zero background $N_{upper}$ is 2.3 at a $90\%$ confidence level. This procedure has been followed to get the upper limit in flux, which is presented in Fig. 7.
\begin{figure}[here]
\center
\includegraphics[width=66mm]{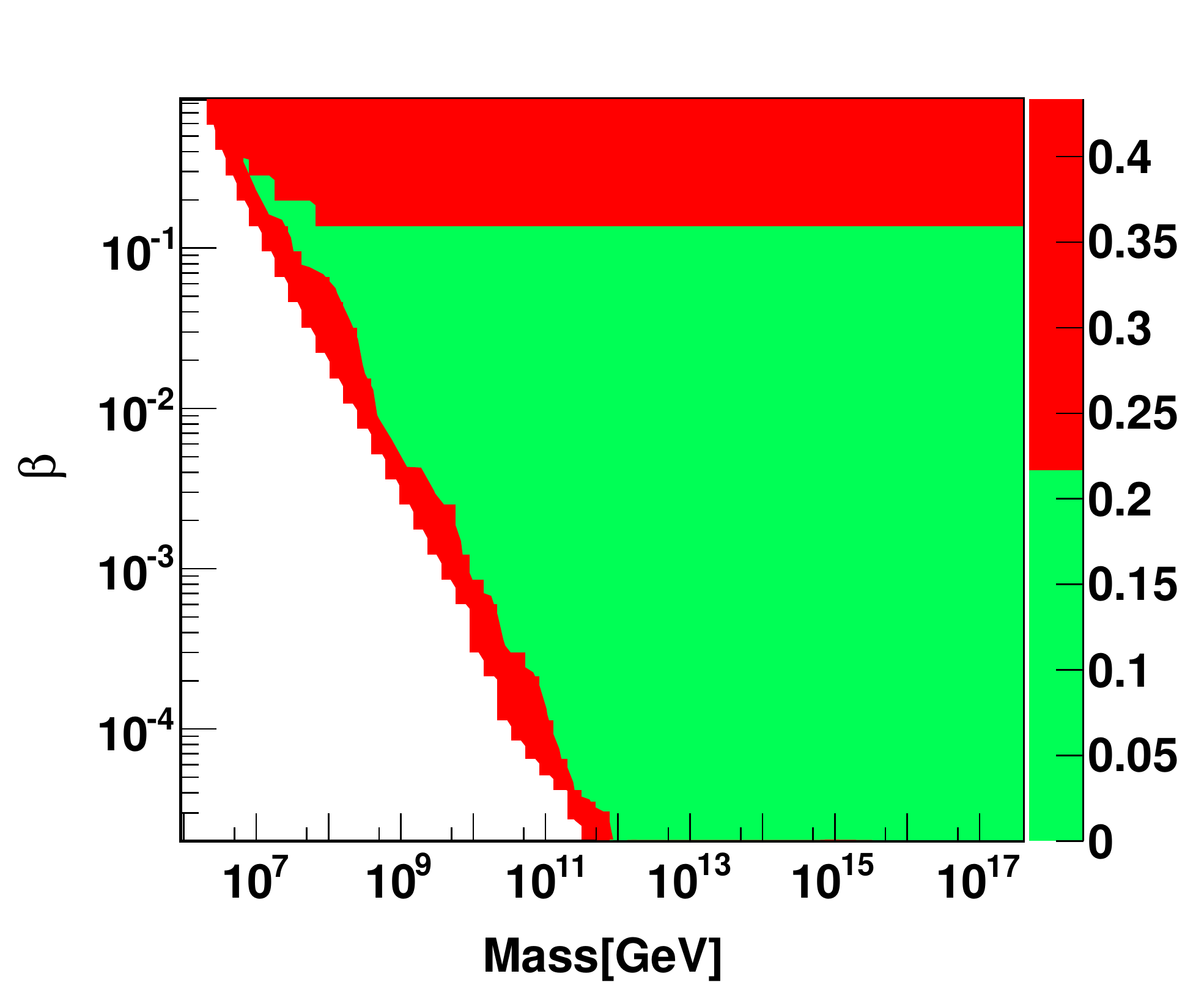}
\caption{\label{fig1}Upper limit in Flux obtained by ICAL with $90\%$ C. L. for 10 years in units of $10^{-15}~cm^{-2}~sr^{-1}~sec^{-1}$.}
\end{figure}

\section{Magnetic Monopole simulation for ICAL engineering module}
An engineering prototype module of the ICAL of dimensions $8~m~\times~8~m~\times~7.5~m~$ is planned to be built over ground at Madurai in the next  $2-3$~years. It has a mass which is 1/8$^{th}$ of that of a single ICAL module. This engineering module will be similar to the main ICAL module from all other points of view. Hence the various parameters of the detector are the same except for the scaling down. The MM simulation has been done to identify the region of M$-\beta$ space where the ICAL prototype has a reasonably good efficiency. 

The same simulation tool-kit has been used as for the main detector module. To simulate the events in the ICAL prototype, the events are generated randomly from the surface of the atmosphere with a height of $10~km$ from the top surface of the prototype detector. The $cos\theta$ is smeared from $\pi~-~\pi/2$ and $\phi$ from $0$ to $2\pi$ to obtained an isotropic flux for MM. For the detection of such type of events on the surface, cosmic ray muons will be the main background. However the sensitivity for lower beta is expected to increase, where one can use the time information to minimise the background due to cosmic muons. Also one has the possibility of covering the lower mass region which is not possible for the underground ICAL due to the energy loss of the MM in around $1~km$ rock cover. For higher $\beta$ ($\beta > 0.2$), the minimum number of layers used for $\beta$ reconstruction is 10 but for lower value of $\beta$ it is used as 5 layers. Figure 8 shows the sensitive region for MM in the M$-\beta$ plane. It covers a region starting from mass $10^5$~GeV to $10^{17}$~GeV and $\beta$ from $10^{-5}$ to $9\times10^{-1}$.
\begin{figure}[here]
\center
\includegraphics[width=66mm]{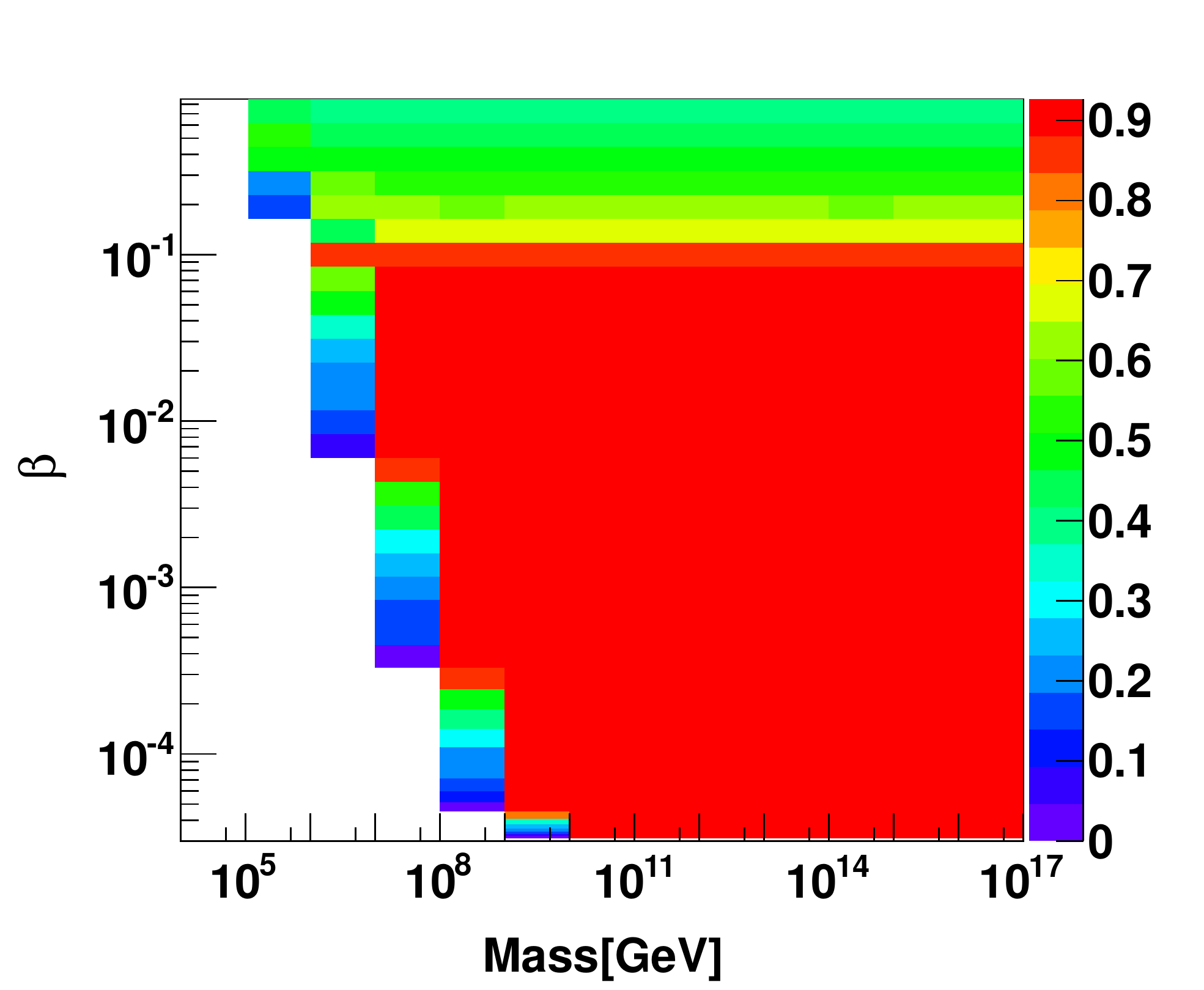} 
\caption{\label{fig1}Efficient region of MM using ICAL prototype in the M$-\beta$ plane. Different colours in the plot presents different values of the efficiencies.}
\end{figure}

In conclusion, using Eq.(5.2), a search for MMs in a certain range of mass and $\beta$ is feasible. This would require a measurement of the time at which the RPC layers fire with time differences between consecutive layers ranging between $0.4~nsec$ and $30$~$\mu sec$, corresponding to $\beta$ of $9\times10^{-1}$ to $10^{-5}$. The flux bounds are comparable to those existing at the present time from experiments worldwide.

\section{Conclusions}
The ICAL at the under ground laboratory can detect MMs with mass ranging from $10^{7}$~GeV to $10^{17}$~GeV. For the mass range of $10^7$ to $10^{9}$~GeV ICAL is sensitive for $\beta \geq 10^{-3}$. For mass from $10^{10}$ to $10^{17}$~GeV ICAL can cover the $\beta$ values above $10^{-5}$. The engineering module of ICAL will cover the lower mass region from $10^5$ - $10^{17}$~GeV. A search for Magnetic Monopoles in a certain range of mass and $\beta$ is feasible with ICAL and also with the ICAL prototype. The flux bounds derived from these detectors will be comparable to those obtained from existing detectors presently operating worldwide.


\section*{Acknowledgements}
The INO project is funded by Department of Atomic Energy (DAE) and Department of Science and Technology (DST) in India. The authors are grateful to Prof M. V. N. Murthy, Prof A. Raychaudhuri and Prof J. B. Singh for their valuable comments and suggestions during the preparation of the manuscript. We would also like to thank Prof A. S. Dighe and P. C. Rout for their worthful suggestions during the analysis. Last, but not least, our thanks are also due to the INO collaborators for their invaluable support.



\begin{thebibliography}{100}
\bibitem{1} http://www.ino.tifr.res.in
\bibitem{2} J. Derkaoui et al., Astroparticle Physics {\bf 9}, 173 (1998).
\bibitem{3} E.N. Parker, Astrophys. J. {\bf 160}, 383 (1970).
\bibitem{4} B. Cabrera, Phys. Rev. Lett. {\bf 48}, 1378 (1982). 
\bibitem{5} S. Bermon et al., Phys. Rev. Lett. {\bf 64}, 839 (1990). 
\bibitem{6} M. Ambrosio et al., [MACRO Collab.], Eur. Phys. J. C {\bf 25}, 511 (2002). 
\bibitem{7} S. Balestra et al., Eur. Phys. J. C {\bf 55}, 57 (2008). 
\bibitem{8} J. L. Thron et al., Phys. Rev., D {\bf 46}, 4846 (1992). 
\bibitem{9} A. Pohl. Search Subrelativistic Particles with the AMANDA Neutrino Telescope. PhD thesis, Uppsala University, 2009.
\bibitem{10} V. Aynutdinov et al., Astropart. Phys. {\bf 29}, 366 (2008).
\bibitem{11} T Kajita et al., J. Phys. Soc. Jap. {\bf 54}, 4065 (1985). 
\bibitem{12} A. Abulencia et al., Phys. Rev. Lett. {\bf 96}, 201801, (2006). 
\bibitem{13} G. R. Kalbfleisch et al., Phys. Rev. Lett. {\bf 85}, 5292 (2000). 
\bibitem{14} P. A. M. Dirac, Proc. Roy. Soc. A {\bf 133}, 60 (1931), P. A. M. Dirac, Phys. Rev. {\bf 74}, 817 (1948).
\bibitem{15} M. N. Saha, Ind. J. Phys. {\bf 10}, 145 (1936)
\bibitem{16} G. 't Hooft, Nucl. Phys. B {\bf 79}, 276 (1974).
\bibitem{17} A. M. Polyakov, JETP Lett. {\bf 20}, 194 (1974).
\bibitem{18} S. P. Ahlen, Phys. Rev. D {\bf 17}, 229 (1978).
\bibitem{19} Y. Kazama et al., Phys. Rev. D {\bf 15}, 2287 (1977).
\bibitem{20} S. P. Ahlen and K. Kinoshita, Phys. Rev. D {\bf 26}, 2347 (1982).
\bibitem{21} S. P. Ahlen and G. Tarle, Phys. Rev. D {\bf 27}, 688 (1983).
\bibitem{22} http://geant4.web.cern.ch/geant4/
\bibitem{23} R. M. Barnett et al., Phys. Rev. D {\bf 54}, 1 (1996).
\end{thebibliography}
\end{document}